# Micromachined piezoelectric membranes with high nominal quality factors in newtonian liquid media: a Lamb's model validation at the microscale

*Cédric Ayela and Liviu Nicu[*]*

LAAS-CNRS, 7, avenue du Colonel Roche, 31077 Toulouse Cedex, France

CORRESPONDING AUTHOR. E-mail: nicu@laas.fr; Phone: +33 5 6133 7838. Fax : +33 5 6133 6208

ABSTRACT. Although extensively presented as one of the most promising silicon-based micromachined sensor adapted to real-time measurements in liquid media, the cantilevered structure still suffers from its quality factor (Q) dramatic dependence on the liquid viscosity thus lowering the measurement resolution. In this paper, micromachined piezoelectric membranes are introduced as a potential alternative to the cantilevers for biological applications. High Q-factors (up to 150) of micromachined piezoelectric membranes resonating in various liquid mixtures (water/glycerol and water/ethanol) are thus reported and a theoretical model proposed by H. Lamb (1920) is validated for microscale structures proving that the variation of the liquid viscosity (if lower than 10cP) has no effect on the dynamic behavior of the membranes. To conclude, two types of experiments were performed in water/glycerol mixtures: in-flow (with liquid continuously flowing on the devices) and in-spot (with individual membranes oscillating in a 5 µL volume of liquid). The results interestingly showed that for the in-spot configuration the Q-factor values are more than twofold the ones corresponding to in-flow measurements thus providing alternative insights into the way to conceive ideal configurations for real-time biological measurements in liquid media.

**I. Introduction**

The biological sensing systems represent nowadays a fast-growing field of research guided by continuously evolving issues as various as the wide range of applications they address, going from common pathologies diagnosis[1] to environmental or biological warfare agents detection[2]. Despite their different targets, the same basic principle is applied to any biosensor: a biorecognition molecule is immobilised over a signal transducer to give a reagentless analytical device. The biorecognition molecule, such as an enzyme, antibody, sequence of DNA, peptide or even a microorganism, provides the biosensor with its selectivity for the target analyte while the signal transducer determines the extent of the biorecognition event and converts it into an electronic signal, which can be outputted to the end user.

Among the common transducers developed up to now, the micro-electromechanical systems seem very promising to meet most of the critical requirements for such specific applications, like the high level of integration for portability purposes, rapid responses, high sensitivity, high signal-to-noise ratio etc. In this class of micro-transducers, silicon-based microcantilevers are the most widespread and proven micromachined biosensors allowing successful characterization of biomolecular systems in their natural aqueous environment[3-5]. When used in mass-sensing configuration, the adsorption of molecules on the surface of a cantilever changes the total mass and, consequently, the resonance frequency of the cantilever.

At this point, it is worth noting that the minimum detectable mass (meaning the smallest amount of target analyte to be detected) is directly dependent on the minimum detectable resonance frequency change, this latter parameter being linked to the quality factor Q of the resonator. In other words, the higher the Q-factor, the better the resolution of the biosensor. When considering that the Q-factor is a measure of the spread of the resonance peak (thus related to energy loss due to damping) it becomes evidence that getting high quality factors in liquid media (specific to biological applications) is one of the most important challenges in the cantilever-based biosensing field.

Until not long ago, Q-factors measured when operating microcantilevers in liquids rarely exceeded 10[6,7]. Several methods have been employed to "artificially" increase this parameter either by feeding back the output signal of the cantilever to the excitation scheme (a.k.a Q-control method)[8], by operating the cantilever on a higher harmonic[9] or by integrating the excitation in the cantilever's structure in order to achieve better actuation efficiency. Even if the first solution cited before allows an improvement of the Q-factor by up to three orders of magnitude, the major drawback is the addition of complementary electronics (phase shifters and gain controlled amplifiers) in the measurement chain. Using higher harmonics thus taking advantage of higher Q-factors is also an elegant method but unfortunately too much dependent on frequency instabilities and coupling.

It remains that the use of an integrated dual excitation-detection scheme (either capacitive or piezoelectric) could be the paved way to obtain higher Q-factors thus avoiding the disadvantages of the methods cited before. If the capacitive excitation is still challenging to adapt for actuation and detection in liquid media[10], the use of piezoelectric thin films (around 1-μm thick) integrated in the resonator's structure have shown promising capabilities in this field since the Itoh et al.[11] pioneering work (although the Q-factor in this latter case was not higher than 8…). More recently, arguing that the forces developed by piezoelectric thin-films are too small to overcome the high effective mass and viscous environment of the liquid, Park et al.[12] proposed the integration of $Pb(Zr_xTi_{1-x})O_3$ (PZT) thick-films (22-μm thick) in a Si-based cantilever structure thus demonstrating strong harmonic oscillations with Q-factors of about 23 in water. Although not exhaustive, the aforementioned examples make evident that the cantilevered microscopic structure is **not intrinsically adapted** to dynamic high-resolution measurements in liquid media because too sensitive to the viscosity of the fluids in which it is immersed.

In this paper, we demonstrate that the use of micromachined circular piezoelectric membranes could be a potential alternative to the cantilevers for biological applications. Although it was extensively shown in the past that this design is well adapted to applications in liquid media (like the piezoelectric micromachined ultrasonic transducers, a.k.a. PMUTs[13]), only few examples make use of it as a potential biosensor[14,15]. To better emphasize this trend, we report high quality factors (up to 200) of micromachined piezoelectric membranes resonating in various liquid mixtures (water/glycerol and water/ethanol) and we experimentally verify the Lamb's model[16] at the microscale, showing that variation of the liquid viscosity (if its absolute value remains lower than 10cP) has no effect on the dynamic behaviour of the devices. The latter result is of crucial importance when thinking "biosensing in liquid media" which viscosity rarely exceeds the value given above. For comparison purpose, blood viscosity at normal hematocrit (volume of blood occupied by red blood cells) rate of 40 is about 3 cP while plasma viscosity (blood from which red cells, white cells and platelets were retired) is about 1.5 cP[17].

## II. Methods and theoretical principles

In this section, the design and the fabrication of the micromachined piezoelectric membranes are discussed, followed by the description of the fluidic experimental set-up specifically developed to validate the Lamb's model at the microscale. Lastly, the hypothesis and fundamentals of the aforementioned theoretical model are briefly reminded.

**Design and fabrication of micromachined piezoelectric membranes**

4x4 matrices of piezoelectric membranes have been fabricated by standard micromachining techniques. The membranes are circular shaped with a total radius (to be called from now $R_2$) equal to 100 µm or 150µm. Each membrane can be individually actuated by a piezoelectric $PbZr_xTi_{1-x}O_3$ (PZT) thin film. This active part is circular shaped with a radius ($R_1$) designed so that three ratios $R_1/R_2$ co-exist in the same matrix (for each $R_2$), respectively equal to 0.3, 0.5 and 0.7 (as depicted in Fig.1). Six different families of membranes have thus been tested in most of the experimental configurations described further.

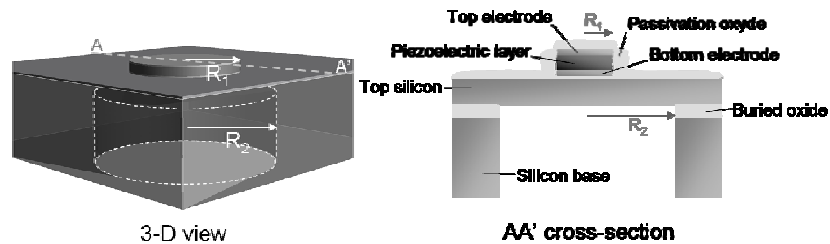

**Figure 1.** 3-D **and cross-section** schemes of a multilayer clamped circular membrane

The main steps of fabrication process are depicted on Fig. 2. The starting substrate is a 100 mm in diameter, <100>, N-type Silicon-On-Insulator (SOI) wafer, with a 1 µm thick buried oxide and a 2 µm thick top silicon layer (resistivity of 7 Ω.cm). The first fabrication step consisted of the growth of 50 nm thick thermal oxide on the entire SOI wafer before the sputtering of Pt/TiO$_2$ (150 nm/ 10 nm) for the bottom electrode. The film was lifted-off with a Shipley 1818 photoresist to define circular electrodes with a $R_1$ radius. The 950 nm thick 54/46 PbZr$_x$Ti$_{1-x}$O$_3$ film was then deposited by RF magnetron sputtering. The 54/46 composition choice is due to its proximity with the morphotropic phase boundary composition (x=53), which has the largest piezoelectric and dielectric constants[18]. In most of studies related to the PZT film deposition, the process temperature is about 650°C. Instead of this, the PZT film used in this study was deposited without intentional substrate heating to allow patterning by lift-off of the PZT film with the Shipley 1818 photoresist[19]. This PZT layer is circular shaped with a radius 5 µm higher than electrode layers to avoid eventual shortcuts between the top and bottom metallic electrodes. Deposited films were therefore amorphous since temperature rise during deposition did not exceed 150°C. After resist stripping, a 30 min crystallization annealing at 625°C was performed. Then, another Ti/Pt (10 nm/ 140 nm) deposition step followed by lift-off was performed in order to obtain top circular electrodes with the same radius as the bottom ones. A passivation silicon oxide film (200 nm) was deposited by Plasma Enhanced Chemical Vapor Deposition (PECVD) in order to avoid electrical contacts between the different active parts of the devices during operation in liquid media. Contact pads were then opened by a wet etching of oxide using HF buffer. To finish, the circular membranes (of $R_2$ global radius) were defined by vertical sidewalls etching on the backside of the SOI wafer using the Deep Reactive Ion etching technique. The 1 µm thick SiO$_2$ acts as an etch stop layer for the dry silicon etching. This layer was then removed using a last Reactive Ion etching step.

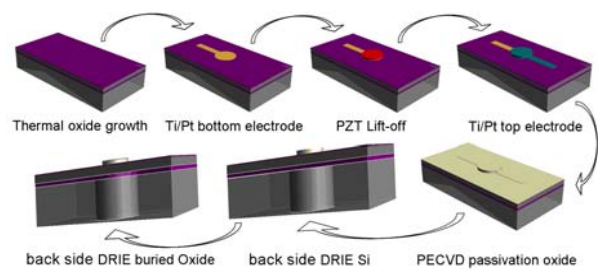

**Figure 2.** Main fabrication steps of micromachined piezoelectric circular membranes

From the packaging point of view, the substrate wafer was diced into 5×5 mm$^2$ individual cells corresponding to one matrix of 16 piezoelectric membranes. Each cell was glued on a TO8 package and each membrane wire-bonded to the TO8 pins. The wires were coated with a biocompatible silicon sealant.

Fig. 3 shows a view of 16 membranes chip and the corresponding silicon-sealed wire bondings. All the membranes are connected with a common ground plane through the bottom electrodes while the top electrode allows individual actuation of the membranes.

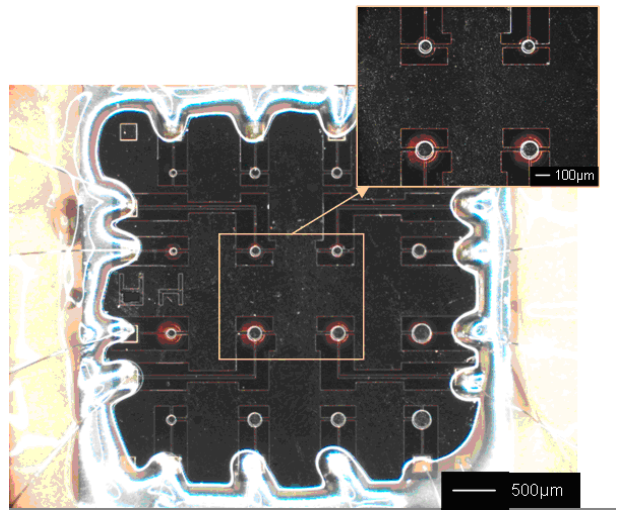

**Figure 3.** Overview of a chip with 16 micromachined circular membranes. Inset shows a close-up view on 4 micromembranes with $R_2$ = 100μm and 150μm and $R_1/R_2$ = 0.5.

**Fluidic experimental set-up**

The fluidic system consists of a flow-through cell, a peristaltic pump and a fluidic circuitry (capillary tubes, adaptive ports, etc.). The flow-through cell (shown in Figure 4) is made of two parts: a top one, made of transparent Plexiglas™ and a bottom one, in aluminium. The top part of the cell was drilled in a depth of 3mm so it fitted exactly on top of the TO8 package. The final distance between the membranes and the bottom of the Plexiglas™ part was estimated at around 2mm. Four plastic screws are holding together the two parts of the fluidic cell and the TO8 package that is placed in-between. The final volume of the cell has been estimated at about 220µL. Two capillary tubes (255/510 µm, Capillary Peek Tubing, Upchurch Scivex) are connected through Nanoport® assemblies (Upchurch Scivex) to the flow-through cell inlet and outlet (1-mm diameter). This set-up allows to the liquid to continuously stream across the piezoelectric membranes with a constant velocity fixed by the peristaltic pump (Masterflex, 1 to 4 mL/h flow rate).

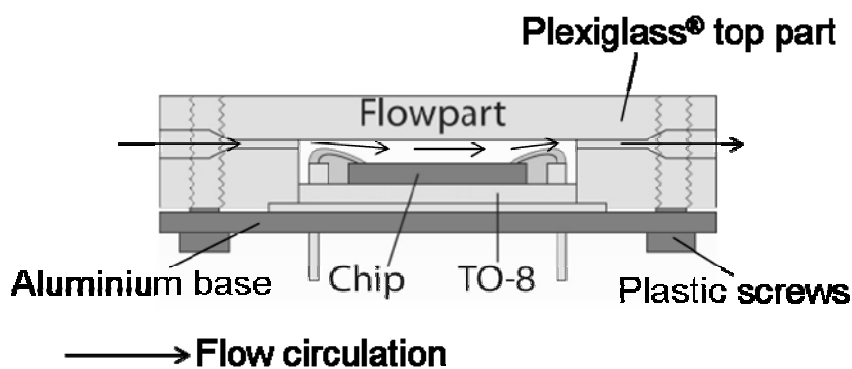

**Figure 4.** Schematics of the flow-through cell (not to scale). The top part is fabricated in Plexiglas™, the base in aluminium and the TO8 package of the piezoelectric membranes matrix is held together with the fluidic cell parts with four plastic screws.

Two kinds of resonant frequency measurements in liquids have been performed: first, liquid was continuously flown onto the membranes in the fluidic cell at a constant flow rate of 4mL/h (called "in-flow experiments") and the resonant frequency was continuously measured. Second, drops of 5µL (with a drop radius of 1.4mm for a measured contact angle of 80°) were deposited onto individual membranes using conventional volumetric pipette and the resonant frequency was measured while the membrane was oscillated into the liquid spot (called "in-spot experiments").

Resonant frequencies of the piezoelectric membranes were determined by performing admittance measurement using a HP4294A Impedance Analyzer from Agilent after poling the PZT film at E = -187 kV/cm (representing 2.5 times the PZT coercive field value) during 20 minutes and using an excitation AC signal of 100 mV in air and 500 mV in liquid. Before plugging the TO8 package to the testing board, an open circuit and short circuit compensation are performed in order to avoid parasitic capacitances due to the connections and the testing set-up.

Calculation of the resonant frequency and the quality factor were done using the Matlab 7.1 software. A linear compensation of the curve due to the static capacitance of the piezoelectric layer was performed before fitting the resonant spectra with a fourth degree polynomial curve. The resonant frequency was identified to the value where the conductance is maximum while the quality factor is calculated by the ratio between the resonant frequency and the frequency bandwidth at $1/\sqrt{2}$ of the conductance's maximum.

**Theoretical principles**

It is a well known fact that as an elastic membrane vibrates in a fluid, the fluid immediately surrounding the vibrating structure is set into motion and thus imposes both added mass and damping forces on the membrane. The added mass will lower the natural frequency of the membrane from that which would be measured in the air while the damping of the membrane will be increased. H. Lamb proposed in 1920 a theoretical model[16] that allows to estimate the shift of the resonant frequency (from vacuum to water, the latter satisfying the assumption of an incompressible liquid) of a thin circular plate filling an aperture in a plane (and rigid) wall which is in contact on one side with an unlimited mass of water.

Starting with an assumed vibration shape of the membrane as follows:

$$w(r,t) = C(t)\left(1 - \frac{r^2}{R^2}\right)^2 \qquad (1)$$

where $w(r,t)$ is the normal displacement of the membrane at a distance $r$ from the centre, $R$ is the membrane radius and $C(t)$ is a function of $t$ only, H. Lamb calculated the kinetic and potential energies of the plate (in vacuum) as being:

$$T_{vac} = \frac{\pi \rho_p h R^2}{10}\left(\frac{dC}{dt}\right)^2$$

$$V_{vac} = \frac{8\pi E h^3}{9(1-\mu)R^2} \qquad (2)$$

where $\rho_p$ is the plate density, $h$ is the plate thickness, $E$ is the value of Young's modulus for the plate material and $\mu$ Poisson's ratio.

In this model, the kinetic energy $T_L$ of the liquid is then calculated and given by:

$$T_L = 0.21 \cdot \rho_L \cdot R^3 \cdot \left(\frac{dC}{dt}\right)^2 \qquad (3)$$

where $\rho_L$ is the liquid density.

Adding $T_L$ to the kinetic energy of the vibrating membrane and applying the Rayleigh-Ritz method (that allows to calculate the resonant frequency of a multi degree of freedom system as a function of the ratio between the maximum values of the potential and the kinetic energies when assuming a known mode shape of vibration) provides an additional factor for the calculation of the resonant frequency in liquid:

$$f_L = \frac{f_{vac}}{\sqrt{1+\beta}} \qquad (4),$$

where $\beta$, usually called added virtual mass incremental (AVMI) factor, is equal to:

$$\beta = 0.67 \cdot \frac{\rho_L R}{\rho_M h} \qquad (5),$$

with $\rho_M$ and $h$ respectively the density and the thickness of the membrane.

At this point, several fundamentals must be outlined: first, Lamb's model does not take into account the viscosity of the liquid medium. One of the main goals of our paper is to prove that this hypothesis is verified if and only if the viscosity value does not exceed around 10cP. Second, as the piezoelectric membranes for experimental validation are multilayered structures (each film of the global stack being subject to internal mechanical stresses), the calculus of the resonant frequency in vacuum is somehow complex and beyond of the scope of this paper. Moreover, we will consider that resonant frequencies in vacuum and in air are identical[20]. This is the reason why $f_{vac}$ values will be implemented in the model from resonant frequency measurements performed in the air. Lastly, Lamb's model also assumes that the membrane's structure is an isotropic monolayer (that is far from the structures studied in this paper).

**III. Experimental results and discussion**

We now examine the dynamic behavior of micromachined piezoelectric membranes by presenting measurements on a range of two distinct families of newtonian fluids with known densities and viscosities. We choose two kinds of mixtures, respectively water/ethanol and water/glycerol which variations of density and viscosity with respect to the proportion of water in the global mixture are radically different. As shown in Fig. 5 (values taken from classical tables), in case of water/glycerol mixture the density and the viscosity values are increasing with the diminution of water proportion in the mixture (with an abrupt rise of the viscosity starting with 60% of glycerol solution). On the contrary, in case of water/ethanol mixture, the density is decreasing with the diminution of water proportion in the mixture while the viscosity is describing a bell-like curve with extrema values between 1cP and 2.5cP.

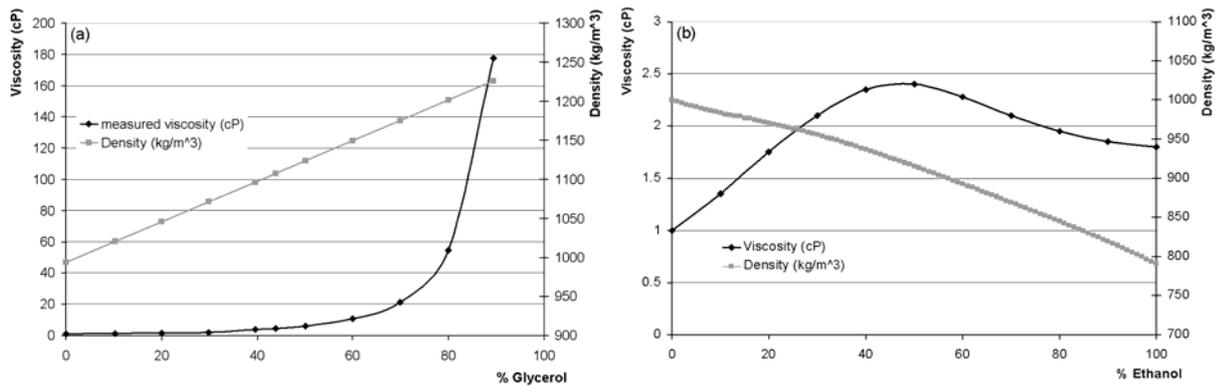

**Figure 5.** Viscosity and density variation vs. (a) glycerol and (b) ethanol proportions in their respective solutions with water

### III.1 Resonant frequency behavior

First resonant frequencies in-flow measurements have been performed in air and in water using six different membranes. The measurements have been limited to the fundamental mode of resonance. An example of resonant frequency spectra in air and in water is shown in Figure 6.

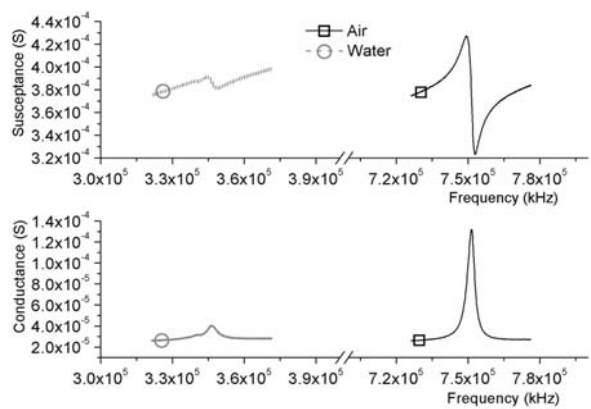

**Figure 6.** Piezoelectric membrane ($R_2 = 100\mu m$, $R_1/R_2 = 0.7$) admittance's imaginary part (susceptance) and real part (conductance) values in air and in water

Lamb's model (eqs 4 and 5) has been used to calculate the theoretical resonant frequencies values in water starting from the measured resonant frequencies in air and using respectively $\rho_{water} = 997 kg/m^3$ for the density of the water and $(\rho h)_{mean}$ instead of $(\rho_M h)$ in eq 5, where $(\rho h)_{mean}$ is the average of the product mass density×thickness of the five-layered composite membrane (the Ti layer has been neglected as it is considerably thinner than the rest of the layers). The theoretical and experimental results are given in Table 1:

| $R_2$ (µm) | $R_1/R_2$ | $f_{0\,air}$ (kHz) | $f_{0\,water}$ Lamb (kHz) | $f_{0\,water}$ experiment (kHz) | Error |
|---|---|---|---|---|---|
| 100 | 0,7 | 751,97 | 346,72 | 346,79 | 0,02% |
| 100 | 0,5 | 625,65 | 262,41 | 286,78 | 8,50% |
| 100 | 0,3 | 580,89 | 208,64 | 237,16 | 12,03% |
| 150 | 0,7 | 270,57 | 89,20 | 89,62 | 0,47% |
| 150 | 0,5 | 295,51 | 85,14 | 94,42 | 9,83% |
| 150 | 0,5 | 337,28 | 97,17 | 103,74 | 6,33% |

**Table 1.** Experimental (air, water) and theoretical (water) values of piezoelectric membranes resonant frequencies measured "in-flow". The last column shows the errors values between measured and calculated resonant frequencies in water.

In Table 1, for membranes with $R_2$ equal to 100µm, the resonant frequency decreases as the piezoelectric layer coverage decreases whereas for membranes with $R_2$ equal to 150µm, the resonant frequency increases. This unexpected resonant frequency dependence might be due to the preponderance of the effective internal stresses effects when the membrane radius is lower than the buckling threshold [21]. On the contrary, in case of a membrane radius value superior to the buckling threshold, the added mass effects might be preponderant.

It should be also noted that besides the fact that Lamb's model is describing well the shift of the resonant frequencies of the piezoelectric membranes at the microscale when moving from air to water, the theoretical values are better fitting to the experience as the ratio $R_1/R_2$ is close to the unity (according to model's assumptions). The latter remark could be explained by the fact the $(\rho h)_{mean}$ calculus is better matching Lamb's model hypothesis concerning the uniform structure of the plate in contact with the water.

The in-flow measurements were continued in water/glycerol mixtures by progressively increasing the proportion of glycerol. Ten different mixtures were prepared containing respectively 10%, 20%...40%, 44%, 50%, 60%...and 90% of glycerol in the global mass of the water/glycerol mixture. Each of them was characterized in terms of viscosity using a commercial AR-G2 rheometer from TA Instruments (the measurements were realized at constant ambient temperature of 22°C). In Figure 7(a-b) we reported respectively the shift of the resonant frequencies from pure water to water/glycerol mixture for two different membranes of global radius $R_2$ respectively equal to 100μm and 150μm (the ratio $R_1/R_2$ being of 0.7 in both cases) and an example of resonant frequency spectra in water and in water/glycerol (80% of glycerol):

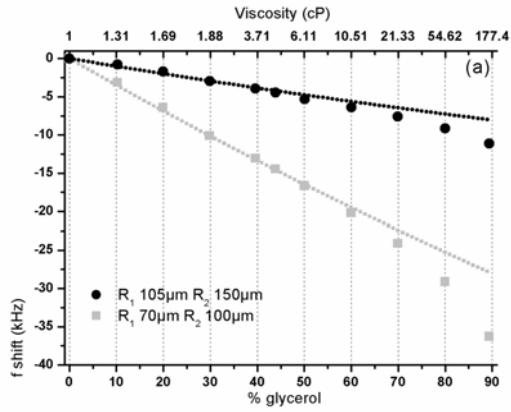 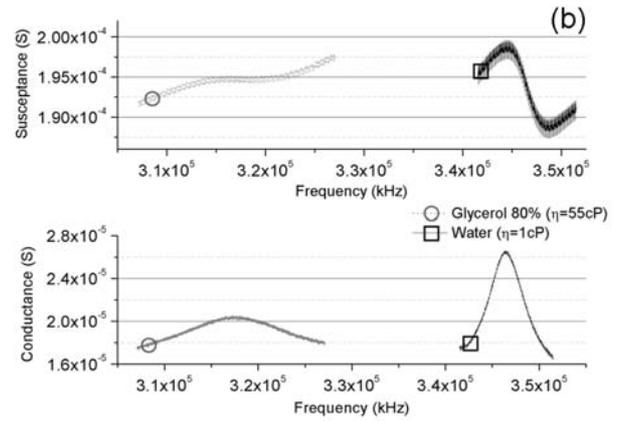

**Figure 7.** (a) Resonant frequencies shift as a function of glycerol content for two piezoelectric membrane designs (symbols stand for experimental values while lines stand for theoretical data); (b) Example of piezoelectric membrane ($R_2$ = 100µm, $R_1/R_2$ = 0.7) admittance's imaginary part (susceptance) and real part (conductance) values in water and in glycerol 80% solution.

Again, the experimental data are well fitting with Lamb's model (continuous line on the graph) until a certain limit that is established around 60% of glycerol in the water/glycerol solution (that is corresponding to a viscosity of 10cP, cf. to Figure 5.a). It should also be pointed out that in the solution containing 80% of glycerol, the resonant frequency is difficult to detect meaning that for solutions with viscosity values higher than 175cP the piezoelectric membranes specifically designed in this work are reaching their detection limit.

In addition, we note that the resonant frequency is decreasing with the increase of the glycerol proportion as the density of the solution is increasing (and so does the AVMI factor). A linear decrease of the resonant frequency with the increase of glycerol might also be observed. This is mainly due to the slight and linear variations of the AVMI factor with the quantity of glycerol allowing, after basic mathematic modifications, the Taylor expansion of equation 4. The first order for the development gives close values compared to the exact function , inducing a linear shift of the resonant frequency with the quantity of glycerol, as shown on figure 7.a. Figure 8 shows the variation of the resonant frequency of one piezoelectric membrane ($R_2$ equal to 100µm, $R_1/R_2$ equal to 0.5) in real time, as a function of time. Each plateau is related to the injection of a different water/glycerol solution. The interrupted line (or the slight discontinuity) on each plateau corresponds to the time during which the piezoelectric membrane was disconnected from the impedance analyzer and the other five membranes (on the same chip) were successively connected in order to measure their resonant frequencies (data previously reported on Figure 7.a). The time duration of the measure was more than 2 hours.

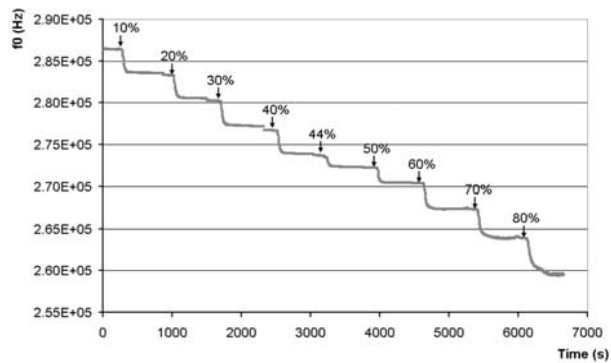

**Figure 8.** Real time measurement of resonant frequency evolution of a piezoelectric membrane ($R_2$ = 100µm, $R_1/R_2$ = 0.5). Transitions between successive plateaus are related to the injection of a different water/glycerol solution.

From these experiments it becomes evident that the viscosity effects on the resonant frequency value can no more be neglected and should be considered in addition to the AVMI factor. In order to validate this hypothesis, we performed the same in-flow measurements in water/ethanol solutions using a 100µm global radius membrane with a $R_1/R_2$ ratio of 0.5. In this case, eight different solutions were prepared containing respectively 10%, 20%...70% and 100% of ethanol in the global mass of the water/ethanol mixture. We reported in Figure 9 the shift of the resonant frequencies of the considered membrane from pure water to water/ethanol mixture.

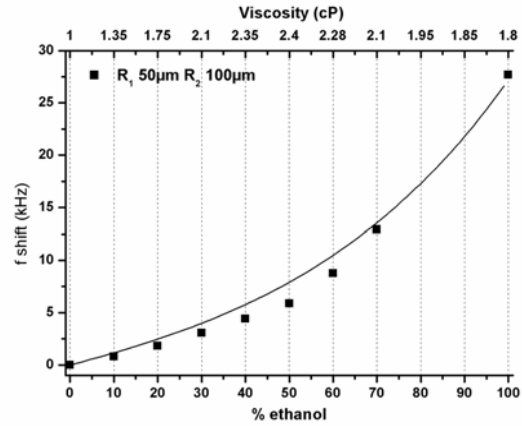

**Figure 9.** Resonant frequency shift as a function of ethanol content for a piezoelectric membrane ($R_2$ = 100µm, $R_1/R_2$ = 0.5). Symbols stand for experimental values while line stands for theoretical data.

As expected, these results are confirming those previously obtained using water/glycerol solutions in that sense that Lamb's model is rigorously verified in case of water/ethanol solutions which viscosities does not exceed 3 cP. In this case, the resonant frequency is varying increasingly with the amount of ethanol thus confirming the fact that the density of the liquid is decreasing at the same time.

**III.2 Quality factor behavior**

The discussion focused so far on the analysis of the resonant frequency shift as a function of the nature of the solution in which the piezoelectric membranes were set into vibratory motion. However, as the effect of the liquid viscosity on the dynamic behavior of the piezoelectric membranes is of primary interest in this study, it would be more interesting to confirm the same behavior for the Q-factor values. In a first experiment, we estimated the Q-factors (defined as the ratio between the frequency value at maximum conductance and full width between 3dB points) from the resonant spectra measured in the previous configuration (in-flow measurements in water/ethanol solutions), as shown in Figure 10:

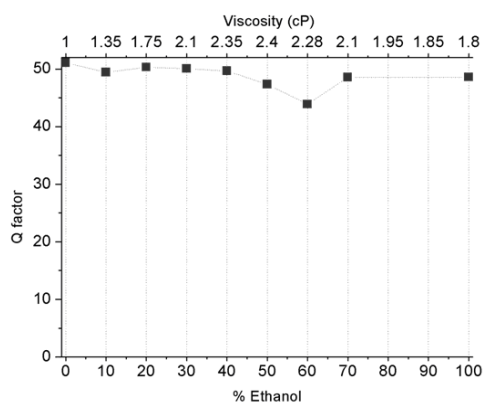

**Figure 10.** Experimental quality factor values of a piezoelectric membrane ($R_2 = 100\mu m$, $R_1/R_2 = 0.5$) versus ethanol proportion in aqueous solutions. For comparison, the Q-factor value in air is about 200.

It can be obviously noted that Q-factors values are almost identical for all water/ethanol solutions tested that confirms again the fact that the effect of the viscosity on the dynamic behavior of the piezoelectric membranes can be neglected in this particular case. Moreover, it also must be outlined that the Q-factors are higher than 50 that is one order of magnitude higher than the Q-factors of most of state-of-the-art micromachined cantilevers used for specific applications in liquid media.

In order to get more insight, we performed Q-factors in-spot and in-flow measurements with water/glycerol mixtures previously prepared. The piezoelectric membrane used here has a global radius $R_2$ equal to 100μm and a $R_1/R_2$ ratio equal to 0.5. In-spot measurements are of primary interest for biological applications in which a drop of analyte (containing the target species) can be spotted onto the active surface of the membrane (previously functionalized with complementary biomolecules) and the resonant frequency monitored during the biological specific recognition phenomenon. The in-spot measurements were impossible to perform in case of water/ethanol solutions because of too high rate of evaporation of the liquid drop spotted on the surface of the chip. As glycerol is well-known as an anti-evaporant agent, it was more suited in the actual configuration. The in-spot and in-flow Q-factors measurements are plotted in Figure 11:

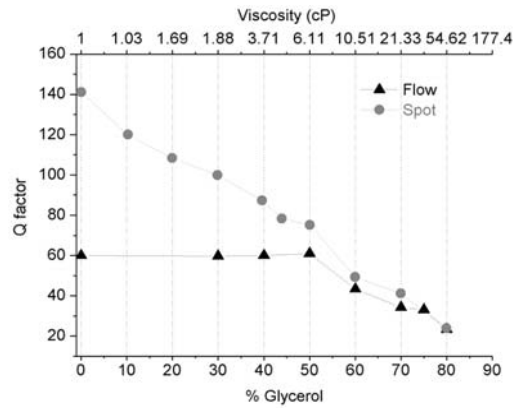

**Figure 11.** In-spot and in-flow measured quality factors of a piezoelectric membrane ($R_2 = 100\mu m$, $R_1/R_2 = 0.5$) versus glycerol proportion in aqueous solutions.

The first remark to be done is that the in-spot Q-factor value is more than twofold the one corresponding to in-flow measurements at the beginning and decreases continuously until it reaches the in-flow value for 60% glycerol solution. One first explanation could be the fact that in the in-spot experiments the liquid drop's mass is sensed by the membrane. In order to raise this doubt, we compared the shift of the resonant frequencies in both configurations. The results are shown in Figure 12:

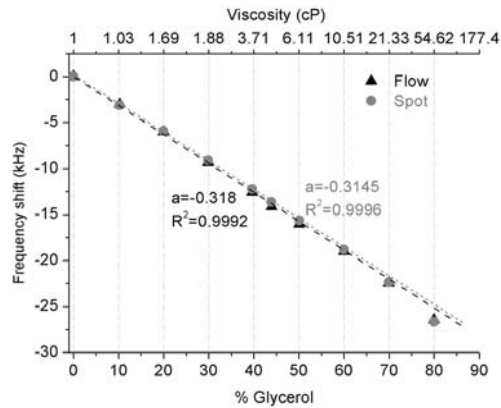

**Figure 12.** In-flow and in-spot comparison of a piezoelectric membrane ($R_2 = 100\mu m$, $R_1/R_2 = 0.5$) resonant frequency shift as a function of glycerol content.

As it can be noted, the frequency shift is rigorously the same for both in-flow and in-spot configurations (as well as the starting frequency values in pure water, in both configurations), thus confirming that in the latter case the liquid drop volume is high enough to be considered as semi-infinite with respect to the membrane dimensions. An eventual interpretation for the high Q-factor value in the in-spot configuration comes from the acoustics theory. When comparing the wavelength of the sound in the liquid to the diameter of the piezoelectric membrane, it appears that in this regime the liquid is mostly reactive meaning that the virtual added mass effects are predominant (compared to the energy loss due to the viscous damping)[22]. Moreover, the compressibility of the interface liquid-air allows the piezoelectric membrane to "freely" oscillate in the liquid drop. On the contrary, in the case of the in-flow configuration, the Q-factor is initially limited (equal to 60) by liquid trapped between the silicon substrate and the rigid wall of the Plexiglass™ fluidic cell. For both configurations, the Q-factors values are decreasing towards the same low values (about 20) as the proportion of glycerol is progressively increased (starting with 60%) in the liquid solution meaning that for high viscosity values the in-spot configuration is no more sufficient to get a higher Q-factor.

To conclude, a last experiment was realized using the in-flow configuration: the experimental conditions were identical to the previous ones; the unique difference consisted in intentionally trapping air bubbles in the fluidic cell (by rapidly reversing the flowing direction of the liquid with the peristaltic pump) near the membranes chip. Q-factors measurements were performed and the results were plotted in Figure 13:

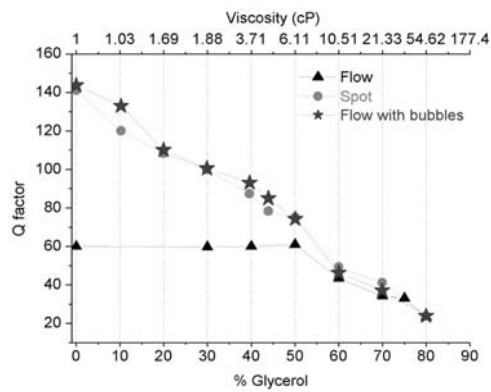

**Figure 13.** In-spot, in-flow and in-flow with air bubbles measured quality factors of a piezoelectric membrane ($R_2 = 100\mu m$, $R_1/R_2 = 0.5$) versus glycerol proportion in aqueous solutions.

Obviously, the Q-factors values of the in-flow configuration (with air bubbles trapped in the fluidic cell) are perfectly matching the in-spot data meaning that the high Q-factors values specific to the latter configuration could be recovered if air bubbles were trapped in a reproducible way in the fluidic cell (though far enough from the active device). This could represent an ideal configuration for biosensing applications where two major requirements have to be imperatively met: continuous flow of the analyte on the surface of the biosensor (in order to permanently refresh the target species and to avoid rebinding phenomena) and high-resolution measurements (realistic only if high Q-factors values are attainable in a liquid medium dynamic measurement configuration).

## IV. Conclusion

In this paper, the use of micromachined circular piezoelectric membranes as a real alternative to the cantilevers for biological applications is demonstrated. To do this, the dynamic behavior in different liquid media of 4x4 matrices of piezoelectric membranes (fabricated by standard micromachining techniques) was studied, each membrane being individually actuated and sensed by a piezoelectric $PbZr_xTi_{1-x}O_3$ (PZT) thin film. Two kinds of resonant frequency measurements have been performed: first, liquid was continuously flown onto the membranes in the fluidic cell at a constant flow rate ("in-flow experiments") and the resonant frequency was continuously measured. Second, drops of 5µl were deposited onto individual membranes and the resonant frequency was measured while the membrane was oscillated into the liquid spot ("in-spot experiments"). The dynamic behavior of the membranes was experimentally investigated on a range of two distinct families of newtonian fluids (water/glycerol and water/ethanol) with known densities and viscosities. Experimental resonant frequency data were in good agreement with theoretical values calculated using the Lamb's model showing that the liquid viscosity effect can be neglected for values lower than 10 cP. For comparison purpose, blood viscosity at normal hematocrit (volume of blood occupied by red blood cells) rate of 40 is about 3 cP while plasma viscosity (blood from which red cells, white cells and platelets were retired) is about 1.5 cP.

Moreover, the piezoelectric membranes exhibited high Q-factors (up to 150) in the aforementioned liquid solutions with values measured in case of the in-spot configuration more than twofold the ones corresponding to in-flow measurements. This last finding suggests that intentionally trapping air bubbles in a fluidic cell containing micromachined membranes for dynamic liquid media measurements could represent an interesting way to improve the measurement capabilities of the sensing devices.

**ACKNOWLEDGMENT**. We gratefully acknowledge Prof. Annie Colin for precious assistance on the water/glycerol mixtures choice and viscosity measurements using conventional rheometry. We also thank Anders Greve for precious assistance on the development of the flow-through cell.

**FIGURE CAPTIONS**

**Figure 1.** 3-D scheme of a multilayer clamped circular membrane

**Figure 2.** Main fabrication steps of micromachined piezoelectric circular membranes

**Figure 3.** Overview of a chip with 16 micromachined circular membranes. Inset shows a close-up view on 4 micromembranes with $R_2 = 100\mu m$ and $150\mu m$ and $R_1/R_2 = 0.5$.

**Figure 4.** Schematics of the flow-through cell. The top part is fabricated in Plexiglas™, the base in aluminium and the TO8 package of the piezoelectric membranes matrix is held together with the fluidic cell parts with four plastic screws.

**Figure 5.** Viscosity and density variation vs. (a) glycerol and (b) ethanol proportions in their respective solutions with water

**Figure 6.** Piezoelectric membrane ($R_2 = 100\mu m$, $R_1/R_2 = 0.7$) admittance's imaginary part (susceptance) and real part (conductance) values in air and in water

**Figure 7.** (a) Resonant frequencies shift as a function of glycerol content for two piezoelectric membrane designs (symbols stand for experimental values while lines stand for theoretical data); (b) Example of piezoelectric membrane ($R_2 = 100\mu m$, $R_1/R_2 = 0.7$) admittance's imaginary part (susceptance) and real part (conductance) values in water and in glycerol 80% solution.

**Figure 8.** Real time measurement of resonant frequency evolution of a piezoelectric membrane ($R_2 = 100\mu m$, $R_1/R_2 = 0.5$). Transitions between successive plateaus are related to the injection of a different water/glycerol solution.

**Figure 9.** Resonant frequency shift as a function of ethanol content for a piezoelectric membrane ($R_2 = 100\mu m$, $R_1/R_2 = 0.5$). Symbols stand for experimental values while line stands for theoretical data

**Figure 10.** Experimental quality factor values of a piezoelectric membrane ($R_2 = 100\mu m$, $R_1/R_2 = 0.5$) versus ethanol proportion in aqueous solutions

**Figure 11.** In-spot and in-flow measured quality factors of a piezoelectric membrane ($R_2 = 100\mu m$, $R_1/R_2 = 0.5$) versus glycerol proportion in aqueous solutions.

**Figure 12.** In-flow and in-spot comparison of a piezoelectric membrane ($R_2 = 100\mu m$, $R_1/R_2 = 0.5$) resonant frequency shift as a function of glycerol content.

**Figure 13.** In-spot, in-flow and in-flow with air bubbles measured quality factors of a piezoelectric membrane ($R_2 = 100\mu m$, $R_1/R_2 = 0.5$) versus glycerol proportion in aqueous solutions.

**TABLES.**

**Table 1.** Experimental (air, water) and theoretical (water) values of piezoelectric membranes resonant frequencies. The last column shows the errors values between measured and calculated resonant frequencies in water.

**REFERENCES**

[1] E. Gizeli and C. R. Lowe, Biomolecular Sensors, Taylor and Francis, London, 2002.

[2] J. J. Gooding, Biosensor technology for detecting biological warfare agents: Recent progress and future trends, Anal. Chim. Acta 559 (2006) 137-151.

[3] J.K. Gimzewski, C. Gerber, E. Meyer and R. R. Schlittler, Observation of a chemical reaction using a micromechanical sensor, Chem. Phys. Lett. 217 (1994) 589-594.

[4] R. Berger, C. Gerber, H. P. Lang and J. K. Gimzewski, Micromechanics: A toolbox for femtoscale science: "Towards a laboratory on a tip", Microelectron. Eng. 35 (1997) 373-379.

[5] G. Abadal, Z. J. Davis, B. Helbo, X. Borris, R. Ruiz, A. Boisen, F. Campabadal, J. Esteve, E. Figueras, F. Perez-Murano and N. Barniol, Electromechanical model of a resonating nano-cantilever-based sensor for high-resolution and high-sensitivity mass detection, Nanotechnology 12 (2001) 100-104.

[6] C. Bergaud and L. Nicu, Viscosity measurements based on experimental investigations of composite cantilevers beam eigenfrequencies in viscous media, Rev. Sci. Instrum. 71 (2000) 2487-2491.

[7] S. Boskovic, J. W. M. Chon, P. Mulvaney and J. E. Sader, Rheological measurements using microcantilevers, J. Rheol. 46 (2002) 891-899.

[8] J. Tamayo, A. D. L. Humphris, A. M. Malloy and M. J. Miles, Chemical sensors and biosensors in liquid environment based on microcantilevers with amplified quality factor, Ultramicroscopy 86 (2001) 167-173.

26**REFERENCES**

[1] E. Gizeli and C. R. Lowe, Biomolecular Sensors, Taylor and Francis, London, 2002.

[2] J. J. Gooding, Biosensor technology for detecting biological warfare agents: Recent progress and future trends, Anal. Chim. Acta 559 (2006) 137-151.

[3] J.K. Gimzewski, C. Gerber, E. Meyer and R. R. Schlittler, Observation of a chemical reaction using a micromechanical sensor, Chem. Phys. Lett. 217 (1994) 589-594.

[4] R. Berger, C. Gerber, H. P. Lang and J. K. Gimzewski, Micromechanics: A toolbox for femtoscale science: "Towards a laboratory on a tip", Microelectron. Eng. 35 (1997) 373-379.

[5] G. Abadal, Z. J. Davis, B. Helbo, X. Borris, R. Ruiz, A. Boisen, F. Campabadal, J. Esteve, E. Figueras, F. Perez-Murano and N. Barniol, Electromechanical model of a resonating nano-cantilever-based sensor for high-resolution and high-sensitivity mass detection, Nanotechnology 12 (2001) 100-104.

[6] C. Bergaud and L. Nicu, Viscosity measurements based on experimental investigations of composite cantilevers beam eigenfrequencies in viscous media, Rev. Sci. Instrum. 71 (2000) 2487-2491.

[7] S. Boskovic, J. W. M. Chon, P. Mulvaney and J. E. Sader, Rheological measurements using microcantilevers, J. Rheol. 46 (2002) 891-899.

[8] J. Tamayo, A. D. L. Humphris, A. M. Malloy and M. J. Miles, Chemical sensors and biosensors in liquid environment based on microcantilevers with amplified quality factor, Ultramicroscopy 86 (2001) 167-173.